\newif\ifepl \epltrue

\ifepl
\documentclass[doublecol]{epl2}
\else
\documentclass[aps,prl,superscriptaddress,twocolumn]{revtex4}
\fi

\usepackage{graphicx,epsf}
\usepackage{amsmath,amssymb}

\newcommand{\w}{\omega}

\newcommand{\TK}{T_{\rm K}}
\newcommand{\Jc}{J_{\rm c}}

\newcommand{\hybb}  {V_0}               
\newcommand{\hyb}   {v}                 
\newcommand{\epsfb} {\varepsilon_0}     
\newcommand{\epsfc} {\varepsilon_c}     
\newcommand{\epsf}  {\varepsilon}       

\ifepl
\newcommand{\onlinecite}{\cite}
\else
\begin{document}
\fi

\title{
Gate-controlled Kondo screening in graphene:\\ Quantum criticality and electron-hole asymmetry
}

\ifepl

\author{Matthias Vojta\inst{1,2} \and Lars Fritz\inst{1} \and Ralf Bulla\inst{1}}
\shortauthor{M. Vojta {\em et al.}}

\institute{
\inst{1} Institut f\"ur Theoretische Physik, Universit\"at zu K\"oln, Z\"ulpicher Str. 77, 50937 K\"oln, Germany \\
\inst{2} Centro Atomico Bariloche, 8400 San Carlos de Bariloche, Argentina
}

\else

\author{Matthias Vojta}
\affiliation{Institut f\"ur Theoretische Physik,
Universit\"at zu K\"oln, Z\"ulpicher Str. 77, 50937 K\"oln, Germany}
\affiliation{Centro Atomico Bariloche, 8400 San Carlos de Bariloche, Argentina}
\author{Lars Fritz}
\author{Ralf Bulla}
\affiliation{Institut f\"ur Theoretische Physik,
Universit\"at zu K\"oln, Z\"ulpicher Str. 77, 50937 K\"oln, Germany}

\fi

\date{April 7, 2010}

\ifepl
\abstract{
Magnetic impurities in neutral graphene provide a realization of the
pseudogap Kondo model, 
which displays a quantum phase transition between phases with screened and
unscreened impurity moment.
Here, we present a detailed study of the pseudogap Kondo model with finite chemical potential $\mu$.
While carrier doping restores conventional Kondo screening at lowest energies,
properties of the quantum critical fixed point turn out to influence the behavior
over a large parameter range.
Most importantly, the Kondo temperature $\TK$ shows an extreme asymmetry between electron and hole doping.
At criticality, depending on the sign of $\mu$, $\TK$ follows either the scaling prediction
$\TK\propto|\mu|$ with a {\em universal} prefactor, or $\TK \propto |\mu|^x$ with $x\approx2.6$.
This asymmetry between electron and hole doping extends well outside the quantum critical
regime and also implies a qualitative difference in the shape of the tunneling spectra
for both signs of $\mu$.
}
\else
\begin{abstract}
Magnetic impurities in neutral graphene provide a realization of the
pseudogap Kondo model, 
which displays a quantum phase transition between phases with screened and
unscreened impurity moment.
Here, we present a detailed study of the pseudogap Kondo model with finite chemical potential $\mu$.
While carrier doping restores conventional Kondo screening at lowest energies,
properties of the quantum critical fixed point turn out to influence the behavior
over a large parameter range.
Most importantly, the Kondo temperature $\TK$ shows an extreme asymmetry between electron and hole doping.
At criticality, depending on the sign of $\mu$, $\TK$ follows either the scaling prediction
$\TK\propto|\mu|$ with a {\em universal} prefactor, or $\TK \propto |\mu|^x$ with $x\approx2.6$.
This asymmetry between electron and hole doping extends well outside the quantum critical
regime and also implies a qualitative difference in the shape of the tunneling spectra
for both signs of $\mu$.
\end{abstract}
\fi


\ifepl
\begin{document}
\fi

\maketitle


\ifepl
\section{Introduction}
\fi
Electrons in graphene provide an essentially perfect realization of two-dimensional (2d)
Dirac fermions \cite{novo1,novo2,neto_rmp}.
Scanning tunneling microscopy (STM) allows to locally study their
electronic structure, including the physics of defects and impurity atoms.
Recently, Kondo resonances for Co atoms adsorbed on heavily doped graphene have been
reported \cite{kondoexp}.
On the theory side, the study of magnetic impurities coupled to 2d Dirac electrons started
in the 1990s in the context of $d$-wave superconductors \cite{withoff,cassa,GBI}.
For a spin-1/2 impurity coupled
to electrons with a density of states (DOS) obeying $\rho(\w) \propto |\w|^r$ -- the
so-called pseudogap Kondo problem (with $r=1$ for $d$-wave superconductors and
graphene) -- it has been established that Kondo screening exists for low temperatures $T$
only if the Kondo coupling $J$ is larger than a critical value $\Jc$. The properties of
the resulting quantum phase transition (QPT) have been studied extensively
\cite{withoff,cassa,GBI,VB01,tolya_SB,VF04,FV04}.
Interestingly, the fixed-point structure changes as function of the bath exponent
$r$ \cite{GBI}.
In particular, for $r > r^\ast \approx 0.375$ the critical
fixed point is particle-hole (p-h) asymmetric, and $r=1$ plays the role of an
upper critical dimension \cite{VF04}.

For an impurity adsorbed on graphene, its position relative to the underlying
carbon honeycomb lattice and the character of its magnetic orbitals determine the electronic
hybridization paths. As a result, orbital degrees of freedom are crucial for Kondo screening
if the impurity is located at a high-symmetry position \cite{achim}.
For the case of Co impurities, the most interesting
situation has been identified as that of a Co atom with spin 1/2 sitting in the center of a C
hexagon. Here, a spin-orbit-driven crossover from SU(4) Kondo physics at high energies to
conventional SU(2) Kondo physics at low energies is predicted. Hence, the low-temperature
behavior is that of a standard Kondo model with a pseudogapped bath DOS.
Moreover, the Kondo coupling $J$ for this model has been estimated to be rather close to
$\Jc$, corresponding to the QPT between screened and unscreened
impurity phases in neutral graphene \cite{achim}.

This opens the exciting possibility to observe Kondo criticality \cite{mvrev} in graphene.
Even if a tuning of the coupling constant exactly to its critical value is impossible,
quantum critical signatures can still be expected at energies above some crossover scale.
The most important new aspect (as compared to $d$-wave superconductors)
is the possibility of doping via a gate voltage, which moves the chemical potential $\mu$
away from the DOS minimum.

\begin{figure}[!t]
\center
\includegraphics[width=2.7in,clip]{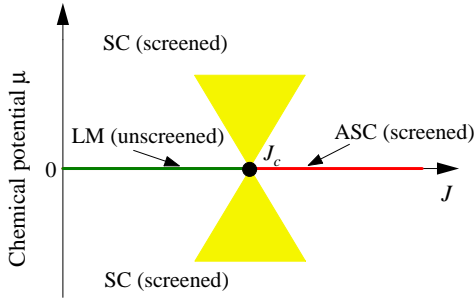}
\caption{
Schematic low-$T$ phase diagram of a pseudogap Kondo impurity
as function of chemical potential $\mu$ and Kondo coupling $J$. LM denotes the local-moment phase
of an unscreened impurity, while SC and ASC are strong-coupling phases with screening.
In the shaded regime (bounded by lines $|\mu|\propto |J-\Jc|^\nu$, with $\nu=1$ for
$r=1$) close to the critical coupling $\Jc$, the universal
prediction $\TK = \kappa_\pm|\mu|$ holds for $r<1$.
At the upper critical dimension $r=1$, this is replaced by
$\TK = \kappa|\mu|$ ($\TK \propto|\mu|^x$) for the two signs of $\mu$.
Strong electron-hole asymmetry of $\TK(\mu)$ is present for a much larger
range of parameters, for details see text.
}
\label{fig:pdsch}
\end{figure}

In this paper, we discuss near-critical Kondo physics in the pseudogap Kondo model with
finite $\mu$, using a combination of analytical and numerical renormalization-group (RG)
methods, with an eye towards possible experiments. As detailed below, we find an
extremely strong asymmetry in the impurity properties between electron and hole doping.
This is less related to the p-h asymmetry of the hybridization functions, but instead is
dictated by the character of the relevant critical fixed point (dubbed ACR) in the
pseudogap Kondo model, which has been shown to have maximal p-h asymmetry near $r=1$
\cite{FV04}. Scale invariance of the critical fixed point, realized for $r<1$, implies
that the Kondo temperature $\TK$ is proportional to the gate voltage,
$\TK=\kappa_\pm|\mu|$, with a universal (and p-h asymmetric) prefactor $\kappa$
\cite{univ_note,troyer}, in a regime close to criticality shown in Fig.~\ref{fig:pdsch}.
Remarkably, we find that logarithmic corrections at the upper critical dimension $r=1$
conspire such that $\TK=\kappa|\mu|$ still holds for one sign of $\mu$, while for the
other sign $\TK\propto|\mu|^x$ with a universal exponent $x\approx2.6$. Even more
importantly, this strong asymmetry in $\TK$ between electron and hole doping extends over
a large parameter range also away from criticality and should thus be easily detectable
in graphene experiments. We also show that impurity tunneling spectra are sensitive to
the sign of $\mu$ as well.

While a number of theory works on Kondo impurities and STM in graphene have appeared
previously \cite{achim,sengupta,zhuang,uchoa,pablo,dellanna},
some of which also discussing aspects of gate-tuned Kondo physics \cite{achim,sengupta},
the dependence of $\TK$ on the gate voltage has not been studied systematically,
and the connection to quantum criticality has not been made precise.
Note here that the often employed methods of poor-man's scaling
and slave bosons provide a qualitatively {\em incorrect} description of the QPT near $r=1$.


\ifepl
\section{Model}
\else
{\it Model.}
\fi
Consider the standard spin-1/2 Kondo model,
$\mathcal{H} = \mathcal{H}_{\rm b} + \mathcal{H}_{\rm imp}$,
with
\begin{equation}
\mathcal{H}_{\rm b} = \sum_{k\sigma} \epsilon_k c_{k\sigma}^\dagger c_{k\sigma},~
\mathcal{H}_{\rm imp} = J {\vec S} \cdot \sum_{\sigma\sigma'} c_{\sigma}(0)^\dagger \frac{\vec\tau_{\sigma\sigma'}}{2} c_{\sigma'}(0)
\label{h}
\end{equation}
with a bath DOS $\rho(\w)$ at site 0 where $c_{\sigma}(0) = N^{-1/2} \sum_k c_{k\sigma}$.
$\vec\tau_{\sigma\sigma'}$ is the vector of Pauli matrices.
Complications arising from orbital degeneracies will be discussed towards the
end of the paper.
For the undoped case, we assume $\rho(\w) = \rho_0 |\w|^r$ for $|\w|$ much smaller than
some crossover scale $\Lambda_P$, while we make no assumptions about $\rho$ for
large energies below the band cutoff $\Lambda$ (where $\rho(\w)$ can be p-h asymmetric).
A finite chemical potential $\mu$ shifts the minimum of $\rho$ away from the Fermi level,
$\rho(\w) = \rho_0 |\w+\mu|^r$, such that $\mu>0$ corresponds to electron doping;
we focus on $|\mu|\ll\Lambda_P$.
For graphene $r=1$, $\Lambda_P \approx 0.5$\,eV \cite{achim} and $\Lambda\approx 8$\,eV.
The low-energy p-h asymmetry of the impurity problem arises from asymmetries of both the
bath and the impurity (the latter coming from additional potential scattering).



\ifepl
\section{Chemical-potential tuned crossover}
\else
{\it Chemical-potential tuned crossover.}
\fi
For $\mu\neq0$, the impurity spin will be screened below a temperature $\TK(\mu)$ for any
$J$. While $J>\Jc$ results in a finite $\TK$ as $\mu\to0$, $J\ll\Jc$ yields an
exponentially small $\TK$, $\ln\TK \propto -1/|\mu|^r$ -- this follows from a
weak-coupling RG treatment (i.e. poor man's scaling). Naively, one would expect only a
weak p-h asymmetry in $\TK(\mu)$ (because of the low-energy p-h symmetry of $\rho(\w)$ at
$\mu=0$) and, in particular, $\TK(\mu)$ to be minimal at $\mu=0$ for $J>\Jc$. Below we
show that these expectations are incorrect.


\ifepl
\section{RG analysis}
\else
{\it RG analysis.}
\fi
We start with the theoretically most interesting case $J=\Jc$.
Scale invariance of the critical system at $\mu=0$
entails $\TK=\kappa|\mu|$ where $\kappa(r)$ is a {\em universal} prefactor
\cite{univ_note}, because $\mu$ scales as an energy and is a relevant perturbation
at the critical fixed point.
We shall show that this scaling prediction is indeed obeyed for $r<1$.
Notably, the prefactor is p-h asymmetric, $\TK=\kappa_\pm|\mu|$ for $\mu\gtrless0$,
for $r^\ast<r<1$ where the critical behavior is controlled by
the ACR fixed point \cite{GBI,FV04}.

The universal critical theory for the QPT near $r=1$, which is not accessible from weak
Kondo coupling, has been worked out in Refs.~\onlinecite{VF04,FV04}.
It is the theory of a crossing of singlet and doublet impurity levels minimally
coupled to conduction electrons, or, equivalently, a $U=\infty$ Anderson impurity model.
Using the notation of Ref.~\onlinecite{FV04}, its impurity part can be written as
\begin{eqnarray}
\label{aiminfu}
\mathcal{H}_{\rm imp} = \epsfb |\sigma\rangle\langle \sigma|
   + \hybb \left[|\sigma\rangle\langle e| c_\sigma(0) + {\rm h.c.}\right]
\end{eqnarray}
where $|\uparrow\rangle$, $|\downarrow\rangle$, and $|{\rm e}\rangle$ represent
the three allowed impurity states. $\epsfb$ is the tuning parameter (``mass'')
of the QPT, i.e., the (bare) energy difference between doublet and singlet states.
The QPT occurs at some $\epsfb=\epsfc$, with screening present for $\epsfb>\epsfc$.
The hybridization $\hybb$ is marginal (relevant) at $r=1$ ($r<1$), i.e.,
and $r=1$ plays the role of an upper critical dimension.
Due to $U=\infty$, the critical fixed point is maximally p-h asymmetric.
Note that the pseudogap Anderson and Kondo models at any fixed $r$ share
the same universality class.

\begin{figure}[!t]
\includegraphics[width=3.3in,clip]{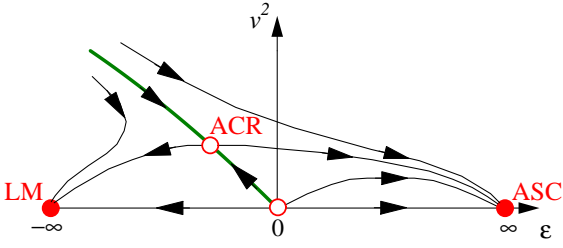}
\caption{
Schematic RG flow \cite{FV04} for the $U=\infty$ Anderson model, \eqref{aiminfu},
for $\mu=0$ and $r^\ast<r<1$.
$\epsf$ and $\hyb$ are the impurity level energy and hybridization, respectively.
The full (open) circles are stable (unstable) fixed points; the thick line separates the
flows to the LM and ASC phases, for details see text.
For $r\to 1$, the ACR fixed points moves towards $\epsf,\hyb=0$.
}
\label{fig:flow}
\end{figure}

Ref.~\onlinecite{FV04} presented a field-theoretic RG analysis of the model
\eqref{aiminfu} for $r\leq 1$, in a double expansion in $\hybb$ and $(1-r)$.
Here we generalize the RG to include a finite $\mu$, restricting
ourselves to the simpler momentum-shell scheme
with ultraviolet (UV) cutoff $D$.
We introduce dimensionless couplings $v$, $\epsf$ according to $\hybb = D^{(1-r)/2} \rho_0^{-1/2} \hyb$
and $\epsfb = D \epsf$ for the hybridization and mass, respectively.
The one-loop RG equations read
\begin{eqnarray}
\frac{dv}{d\ln D} &=& -\frac{1-r}{2} \hyb + \frac{\hyb^3}{2} F_1(\mu/D), \nonumber \\
\frac{d \epsf}{d\ln D} &=& -\epsf + \hyb^2\epsf \, F_1(\mu/D) + \hyb^2 \,F_2(\mu/D)
\label{rg}
\end{eqnarray}
with $F_{1,2}(y) = |1+y|^r \pm 2 |1-y|^r$.
The last term in $d\epsf/d\ln D$ simply
describes the level shift due to the real part of the bath Green's function. For $\mu=0$,
the RG equations \eqref{rg} reduce to those of Ref.~\onlinecite{FV04},
with the flow depicted in Fig.~\ref{fig:flow}.
They yield a critical fixed point at ${\hyb^\ast}^2 = (1-r)/3$ and $\epsf^\ast = -(1-r)/(3r)$ for
$r\lesssim 1$ -- this is the ACR fixed point. The fixed points $\epsf=\pm\infty$
correspond to the screened and unscreened phases, ASC and LM, respectively.

For small $\mu\neq0$, the flow at $\epsfb=\epsfc$ deviates from the critical fixed point on the
scale $|\mu|$ \cite{frgnote}.
For negative $\mu$, the flow is directly into the screened (ASC) phase, $\epsf\to\infty$,
implying $\TK\sim|\mu|$, i.e., $\kappa_- = \mathcal{O}(1)$.
For a more detailed analysis, we can deduce $\TK$ from the flow of $m\equiv\epsf/\hyb^2$.
From Eqs.~\eqref{rg} one finds
\begin{eqnarray}
\label{rg2}
\frac{dm}{d\ln D} = -m r + F_2(\mu/D)
\end{eqnarray}
with the critical-point value $m^\ast=-1/r$.
For $\mu\neq 0$, we apply a two-stage RG procedure \cite{frgnote}:
first, the cutoff $D$ is reduced down to $\mu$, following the flow equations for
$\mu=0$. The second stage starts at $D=\mu$ (and $m=m^\ast$ at $J=J_c$),
and follows Eqs.~(\ref{rg},\ref{rg2}).
For simplicity, we may approximate here $F_{2}(y) \approx - |y|^r$,
which allows to integrate Eq.~\eqref{rg2} analytically.
Finally, $\TK$ may be defined as the value of $D$ where $m$ has changed by unity,
$m(\TK)=m^\ast+1$, which yields
\begin{equation}
\label{plog}
\TK = |\mu| [W(\{r-1\}/e) / (r-1)]^{1/r}
\end{equation}
where $W(x)$ is the Lambert $W$ (or product log) function and $e=2.71828\ldots$

In contrast, for small positive $\mu$ the flow at $\epsilon_c$ is driven towards negative
$\epsf$, i.e., the doublet (LM) fixed point, which implies that a spin 1/2 degree of freedom
is formed on the scale $|\mu|$. This moment is subsequently screened due to the finite
DOS at the Fermi level -- this second crossover is akin to conventional Kondo screening.
It involves restoration of p-h symmetry and is therefore not described by the RG
equations \eqref{rg}. To obtain a rough estimate of $\TK$ here, we may again envision
a second stage of RG, which starts at $D=\mu$ and now uses the language of poor man's scaling. This
results in a conventional exponential expression for $\TK$, here to be used with
bandwidth $\mu$ and dimensionless hybridization $\sim\sqrt{1-r}$ (taken from the
fixed-point value $\hyb^\ast$). Consequently, $\TK=\kappa_+|\mu|$, with $\ln\kappa_+
\propto -1/(1-r)$, i.e., $\TK\ll|\mu|$ for $r\lesssim 1$.

For $r=1$, the upper critical dimension of the problem, one may expect logarithmic
corrections due to the logarithmic critical flow towards $\epsf = \hyb=0$, e.g.,
$\hyb^2 = \hybb^2 / [ 1-3\hybb^2\ln(D/D_0) ]$ where $D_0$ is the
initial UV cutoff \cite{FV04}.
Remarkably, $m^\ast = -1/r$ also applies to this logarithmic flow,
and hence logarithmic corrections are {\em absent} from the $\TK$ expression
\eqref{plog} for $\mu<0$ \cite{lognote}.
For $\mu>0$, the above two-stage RG analysis now suggests $\ln\kappa_+ \propto
\ln(\mu/D_0)$ because the relevant dimensionless hybridization is $v^2 \propto -1/\ln(\mu/D_0)$.
Hence, $\kappa_+$ becomes scale-dependent:
the scaling prediction is spoiled and replaced by $\TK/D_0 \propto |\mu/D_0|^x$
where $x>1$ is now a {\it universal} exponent.
(To obtain an estimate for $x$ here, a detailed analysis of the crossover between the two
RG stages would be required which we do not attempt.)
$\TK\propto|\mu|^x$ also implies that the low-energy impurity physics for $\mu>0$ is
characterized by two distinct scales, $\mu$ and $\TK$.

Let us summarize the main physics at this point.
The fundamental asymmetry between electron and hole doping arises from the p-h asymmetric
character of the critical theory \eqref{aiminfu}. Its phase transition can be driven by
varying the real part of the bath Green's function.
Now, varying $\mu$ leads to a {\em linear} variation
of this real part, and hence drives the critical system either towards the singlet or the
doublet phase on the scale $|\mu|$.
We naturally expect that the electron-hole asymmetry, existing at
the critical coupling, pertains to the off-critical situation as well. Our numerical
results (Fig.~\ref{fig:colltk}) show that this is indeed the case.

For completeness, we also mention the result for $\TK(\mu)$ at criticality for small $r$.
Here, the slave-boson mean-field treatment of \eqref{h}
\cite{withoff} gives qualitatively correct results for the p-h symmetric critical (SCR) fixed
point \cite{GBI,FV04}.
The mean-field equation determining $\TK$ can be written as
\begin{equation}
\frac{1}{J} = \int d\w \frac{\rho(\w)}{\w} \tanh\frac{\w}{2\TK}\,.
\end{equation}
At $\mu=0$, setting $\TK=0$ gives the critical coupling $\Jc$.
For a system with $\rho(\w) = \rho_0|\w\pm\mu|^r$ and $J=\Jc$,
$\TK(\mu)$ follows from the dimensionless equation
\begin{equation}
\int dx \left(|x|^{r-1} - \frac{|x\pm 1|^r}{x} \tanh\frac{x}{2\bar{\TK}} \right) = 0
\label{tkeq}
\end{equation}
where $x=\w/|\mu|$, $\bar{\TK}=\TK/|\mu|$.
For $r<1$, the ultraviolet cutoff of the integral can be sent to infinity.
Note that the non-universal factor $\rho_0$ has dropped out.
The result for $\TK$ is independent of the sign of $\mu$,
$\TK = \kappa |\mu|$.
Solving Eq.~\eqref{tkeq} yields $\kappa\propto\exp(-1/r)$ for small $r$.
The same result can be obtained using a two-stage version of poor man's scaling.
In the first stage, $\mu=0$, the RG equation for the dimensionless Kondo coupling $j$
is $dj/d\ln D = -rj + j^2$ \cite{FV04}. In the second RG stage, starting at the scale $\mu$,
the fixed-point coupling $j^\ast=r$ simply enters the conventional exponential expression
for $\TK$, i.e., $\TK = |\mu| \exp(-1/r)$.


\begin{figure}[!t]
\includegraphics[width=3.3in,clip]{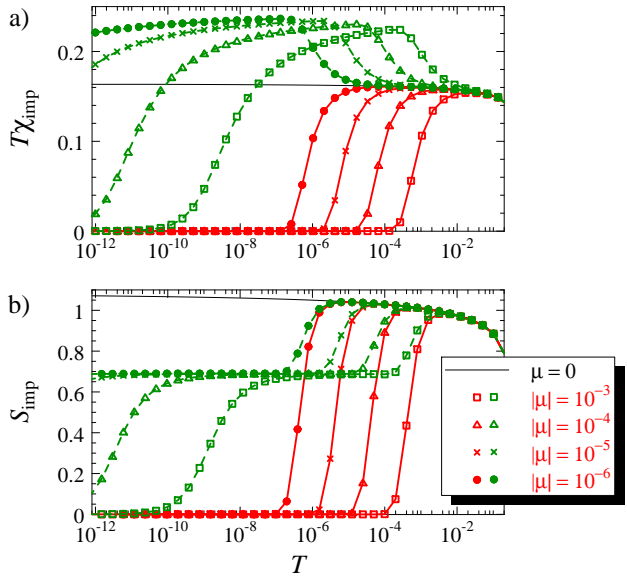}
\caption{
NRG results \cite{NRG_para} for impurity moment and entropy, $T\chi_{\rm imp}$ and $S_{\rm imp}$,
for a $U=\infty$ Anderson model \eqref{aiminfu} at the critical point, $\epsfc=-0.3105935$,
for different $\mu$, with dashed (solid) curves for $\mu>0$ ($\mu<0)$.
The bath DOS is $\rho(\w)=|\w+\mu|\Theta(1-|\w+\mu|)$, i.e., $r=1$, and the hybridization $V_0^2=1/\pi$.
The screening process displays an extreme p-h asymmetry.
}
\label{fig:xover}
\end{figure}

\ifepl
\section{NRG results}
\else
{\it NRG results.}
\fi
All the above considerations are well borne out by simulations using
Wilson's numerical RG (NRG) technique \cite{nrgrev}.
In Fig.~\ref{fig:xover} we display NRG results for the impurity magnetic moment $T\chi_{\rm imp}$
and entropy $S_{\rm imp}$ of a $U=\infty$ Anderson model with symmetric $r=1$ DOS
at the critical coupling, $\epsf=\epsfc$, for different $\mu$.
At the critical fixed point $T\chi_{\rm imp} = 1/6$ and $S_{\rm imp}=\ln 3$ \cite{VF04}.
Various features are apparent in Fig.~\ref{fig:xover}:
The observables deviate from their critical value for $T\lesssim|\mu|$.
More importantly, there is a striking asymmetry between the two signs
of $\mu$, e.g., the magnetic moment $T\chi$ increases (decreases) for $\mu>0$ ($\mu<0$)
once $T$ drops below $|\mu|$. This reflects the fact that the system is driven towards the
doublet or singlet phase depending on the sign of $\mu$.
Further, for $\mu>0$ a clear two-stage crossover is seen, where a spin-1/2 moment is
formed below $T\sim|\mu|$, with $T\chi_{\rm imp}\sim 1/4$ and $S_{\rm imp}\sim\ln 2$, which is only screened at
very low $T$. The power-law behavior, $\TK(\mu) \propto \mu^x$ for $\mu>0$,
is visible, and a fit gives $x\approx 2.6\pm 0.1$.

\begin{figure}[!t]
\includegraphics[width=3.45in,clip]{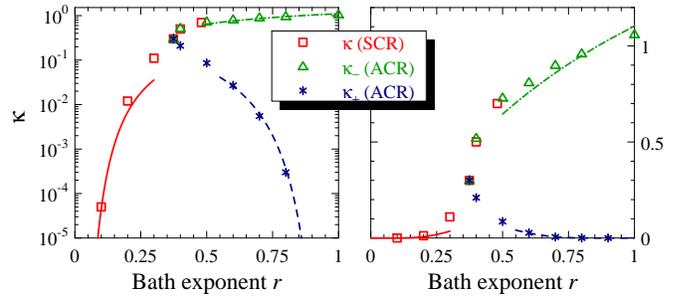}
\caption{
Universal prefactors $\kappa$ of $\TK = \kappa_\pm|\mu|$, obtained from NRG \cite{NRG_para},
for the symmetric (SCR) and asymmetric (ACR) critical fixed points.
Both panels show the same data, with log and linear $\kappa$ axes, respectively.
$\TK$ is defined here via thermodynamic observables \cite{tkdef}.
The lines represent the asymptotic forms
$\exp(-1/r)$ (solid),
$1.8\exp[-1.7/(1-r)]$ (dashed), and
$3[W(\{r-1\}/e) / (r-1)]^{1/r}$ (Eq.~\ref{plog}, dash-dot),
for details see text.
}
\label{fig:kappa}
\end{figure}

\begin{figure}[!b]
\includegraphics[width=3.4in,clip]{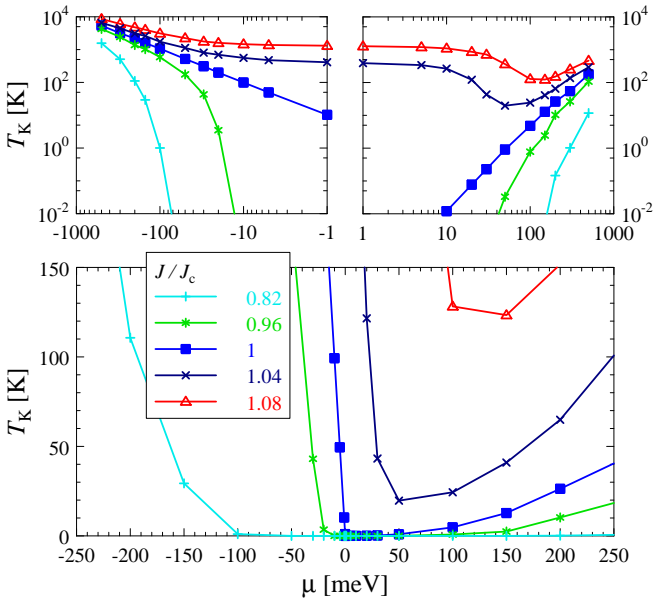}
\caption{
NRG results for the Kondo temperature as function of $\mu$ for different values
of the Kondo coupling $J$, calculated for a DOS appropriate for Co on graphene \cite{dosnote}
where $\Jc \approx 4.3$\,eV \cite{jconv}.
Both top and bottom panels show the same data, with log and linear axes, respectively.
}
\label{fig:colltk}
\end{figure}

Fig.~\ref{fig:kappa} shows the universal prefactors $\kappa(r)$, obtained from NRG,
of the ``critical'' relation $\TK=\kappa_\pm|\mu|$.
Note that $\TK$ has been extracted from $T\chi_{\rm imp}$ \cite{tkdef};
in general, $\TK$ is only defined up to a prefactor of order unity.
We have verified universality of $\kappa$ by performing calculations for
both Kondo and $U=\infty$ Anderson models and for various power-law DOS with different
high-energy parts.
The results for $\kappa$ nicely follow the asymptotic forms derived above,
i.e., appear well fitted by
$\kappa(r\to0)   \approx \exp(-1/r)$,
$\kappa_+(r\to1) \propto \exp[-1.7/(1-r)]$, and
$\kappa_-(r\to1) \propto [W(1+r)/(1+r)]^{1/r}$.
NRG results away from the critical coupling will be discussed in connection
with graphene in the next section.

We finally note that the role of the two signs of $\mu$ is interchanged if the sign of
the model's p-h asymmetry is switched. Our quoted signs are valid for a $U=\infty$
Anderson model, and also apply for a Co impurity on graphene, assuming a Kondo model and
an orbital E1 configuration of Co \cite{achim}, see below.

\ifepl
\section{Application to graphene}
\else
{\it Application to graphene.}
\fi
For Co in the center of a graphene's hexagon, the hybridization function is strongly
asymmetric on the scale of a few eV \cite{achim}. We have used a $\rho(\w)$ which
resembles \cite{dosnote} the tight-binding fit to the LDA DOS in Fig.~2 of
Ref.~\onlinecite{achim} (for Co in E1 configuration)
to calculate $\TK$ in a standard Kondo model as function of Kondo coupling $J$ and chemical potential $\mu$.
Results are displayed in Fig.~\ref{fig:colltk}.
The strong electron-hole asymmetry is apparent, which continues to exist away from the
critical coupling.
At $J=\Jc$ the linear and power-law behaviors of $\TK(\mu)$ are nicely visible.
Moreover, the minimum of $\TK(\mu)$ for $J>\Jc$ is found at significant positive bias.

A few comments are in order:
(i) For simplicity, we have assumed a simple SU(2) spin 1/2 impurity.
According to Ref.~\onlinecite{achim}, spin-orbit coupling reduces the symmetry
of the Co impurity problem from SU(4) to SU(2) on a scale of 60 meV. This implies that
our results are experimentally relevant only below a temperature of $\sim 600$\,K \cite{flownote}.
Moreover, the Kondo coupling in our SU(2) model has to be taken essentially
by a factor of 2 larger compared to the bare Kondo coupling of the Co impurity to obtain
the same $\TK$ (because the impurity degeneracy multiplies the Kondo coupling within
the poor man's scaling equation.)
Hence, we predict a $\Jc$ of 2.2\,eV for Co which approximately matches \cite{jconv}
the estimate in Ref.~\onlinecite{achim}.
(ii) The existing STM data, with $\TK\approx 15$\,K at $|\mu|=0.2$\,eV
\cite{kondoexp}, can either be consistent with our results for $\mu<0$ and $J$
significantly smaller than $\Jc$ or with $\mu>0$ and $J\approx\Jc$.
The two cases differ in the change of $\TK$ upon inverting the sign of $\mu$:
in the former case $\TK$ should become vanishingly small, but huge in the latter
(which would still be measurable if $\mu$ is reduced).
(iii) The results in Fig.~\ref{fig:colltk} are switched according to
$\mu \leftrightarrow -\mu$ if the Co impurity is in an E2 configuration, because
the hybridization functions for E1 and E2 symmetry are approximately identical after p-h
transformation \cite{achim}.

\begin{figure}[!t]
\includegraphics[width=3.4in,clip]{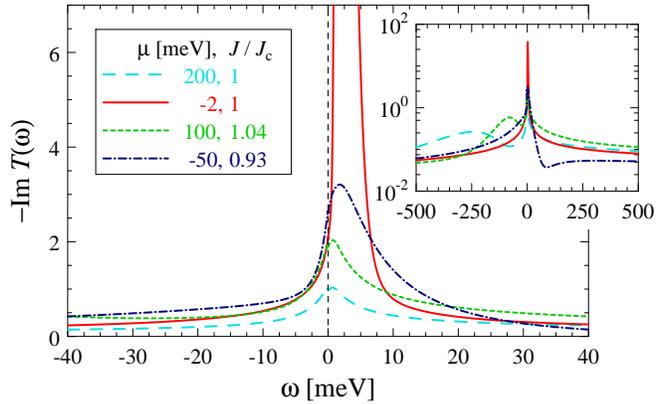}
\caption{
$T$ matrix spectral function for a Kondo impurity coupled to a DOS
describing Co on graphene \cite{dosnote}, for different values of the chemical potential and the Kondo
coupling. All cases have a $\TK$ between 20 and 30\,K, but very different spectral
shapes. The inset shows the same data on a larger scale. Note that the $\mu>0$ spectra
have a broad peak around $(-\mu)$ in addition to the Kondo peak.
}
\label{fig:spec}
\end{figure}

\ifepl
\section{Tunneling spectra}
\else
{\it Tunneling spectra.}
\fi
The different screening processes for $\mu\gtrless0$ result in distinct
spectral functions as well. In Fig.~\ref{fig:spec} we show NRG results for the impurity
$T$ matrix at $T=0$, for situations with similar $\TK$, but reached with different combinations of
$\mu$ and $J$.
Most striking is again the case $J=\Jc$, where
the high-energy part of the spectrum follows the critical law
${\rm Im} T(\w) \propto 1/(\omega |\ln\omega|^2)$ \cite{cassa,FV04}.
For $\mu<0$, the RG flow from critical to
singlet behavior results in a large peak in ${\rm Im}T(\w)$ of width $\TK$ {\em away} from the Fermi level,
located at $\w\sim\TK\sim -\mu$.
In contrast, the two-stage screening for $\mu>0$ yields a more conventional Kondo peak
{\em at} the Fermi level, but in addition a broad peak at a much
larger energy $\w\sim -\mu$ (see inset of Fig.~\ref{fig:spec}).
This behavior persists away from $J=\Jc$, i.e., the Kondo peak has a pronounced asymmetry
w.r.t. the Fermi level for $\mu<0$ (and $J$ not too far from $\Jc$).

The impurity $T$ matrix spectrum, Fig.~\ref{fig:spec}, is in general not identical
to the signal in a STM experiment,
due to additional tunneling contributions into the host orbitals. The interference between
the two is known to convert Kondo peaks into Fano lineshapes, with the Fano parameter
depending upon details of the tunneling process \cite{fanoth}.
Nevertheless, the position and width of the Kondo feature are independent of this interference.
Hence, we predict that for $J\sim\Jc$ the Kondo feature in STM will either be centered
at the Fermi level or be located at an energy of order $\TK$ away from the Fermi level with a
sign opposite to $\mu$, depending on the sign of $\mu$.

\ifepl
\section{Conclusions}
\else
{\it Conclusions.}
\fi
We have discussed Kondo screening for magnetic impurities in doped pseudogap Fermi
systems, with focus on the effects of nearby quantum criticality.
For bath exponents $r<1$, we found that the Kondo temperature follows
the universal law $\TK=\kappa_\pm|\mu|$ for $\mu\gtrless0$ in the critical regime.
For $r=1$, a subtle interplay of p-h
asymmetry and logarithmic corrections at the upper critical dimension leads to an extreme
p-h asymmetry in $\TK(\mu)$, with linear or power-law behavior being
realized depending on the sign of $\mu$.

For a situation corresponding to Co in graphene, we have predicted both the dependence of
$\TK$ on $\mu$ and characteristics of low-temperature STM spectra. Most striking is the
strong asymmetry between electron and hole doping over a large parameter range. This
behavior is surprising because the relevant graphene DOS is weakly p-h asymmetric at low
energies and hence a strong asymmetry is not expected from standard weak-coupling
approaches. Instead, it is the strong p-h asymmetry of the critical fixed point of the
pseudogap Kondo model which is responsible for the physics advertised here
(and which is not captured by slave-boson or weak-coupling approaches to the Kondo
problem).

Our findings call for systematic measurements of $\TK$ as function of $\mu$, in
particular for gate voltages in the linear DOS regime, i.e., below 0.2\,eV. The asymmetry
of $\TK(\mu)$, together with the ab-initio data of Ref.~\onlinecite{achim}, will allow
to determine the orbital nature of the Co impurity. Moreover, a linear or power-law behavior
of $\TK(\mu)$ will be a strong indication of nearby quantum criticality at $\mu=0$.



We are grateful to P. Cornaglia, G. Usaj, and in particular A. Rosch
for discussions.
This research was supported by the DFG through SFB 608, SFB-TR 12 and FG 960.
MV also acknowledges support by the Heinrich-Hertz-Stiftung NRW and the
hospitality of the Centro Atomico Bariloche where part of this work was performed.



\begin{thebibliography}{99}

\bibitem{novo1}
K. S. Novoselov {\em et al.},
Science {\bf 306}, 666 (2004).

\bibitem{novo2}
K. S. Novoselov {\em et al.},
Nature {\bf 438}, 197 (2005).


\bibitem{neto_rmp}
A. H. Castro Neto {\em et al.},
Rev. Mod. Phys. {\bf 81}, 109 (2009).

\bibitem{kondoexp}
L. S. Mattos {\em et al.}, to be published.

\bibitem{withoff}
D.~Withoff and E.~Fradkin, Phys. Rev. Lett. {\bf 64}, 1835 (1990).

\bibitem{cassa}
C. R. Cassanello and E. Fradkin, Phys. Rev. B {\bf 53}, 15079 (1996).

\bibitem{GBI}
C.~Gonzalez-Buxton and K.~Ingersent, Phys. Rev. B {\bf 57}, 14254 (1998).

\bibitem{VB01}
M. Vojta and R. Bulla, Phys. Rev. B {\bf 65}, 014511 (2001).

\bibitem{tolya_SB}
A. Polkovnikov, Phys. Rev. B {\bf 65}, 064503 (2002).

\bibitem{VF04}
M. Vojta and L. Fritz, Phys. Rev. B {\bf 70}, 094502 (2004).

\bibitem{FV04}
L. Fritz and M. Vojta, Phys. Rev. B {\bf 70}, 214427 (2004).

\bibitem{achim}
T. O. Wehling {\em et al.},
Phys. Rev. B {\bf 81}, 115427 (2010).

\bibitem{mvrev}
M. Vojta, Phil. Mag. {\bf 86}, 1807 (2006).

\bibitem{troyer}
A universal perturbation-induced transition temperature near quantum criticality,
similar in spirit to our universal $\TK=\kappa_\pm|\mu|$,
has been discussed for bilayer magnets in:
M. Troyer and S. Sachdev,
Phys. Rev. Lett. {\bf 81}, 5418 (1998).

\bibitem{univ_note}
{\em Universal} means that $\kappa$ does only depend on the bath DOS exponent $r$,
but neither on its prefactor $\rho_0$ nor on high-energy properties of the DOS.

\bibitem{sengupta}
K. Sengupta and G. Baskaran,
Phys. Rev. B {\bf 77}, 045417 (2008).

\bibitem{zhuang}
H.-B. Zhuang, Q.-f. Sun, and X. C. Xie,
EPL {\bf 86}, 58004 (2009).

\bibitem{pablo}
P. S. Cornaglia, G. Usaj, and C. A. Balseiro,
Phys. Rev. Lett. {\bf 102}, 046801 (2009).

\bibitem{uchoa}
B. Uchoa {\em et al.},
Phys. Rev. Lett. {\bf 103}, 206804 (2009).

\bibitem{dellanna}
L. Dell'Anna,
J. Stat. Mech. P01007 (2010).

\bibitem{frgnote}
For $\mu\neq0$, the problem at hand is not scale-invariant.
This renders the standard coupling-constant RG insufficient for quantitative
calculations, because all vertex functions will develop structures on the scale $\mu$.
Our two-stage RG scheme partially accounts for this complication;
a more accurate treatment could be achieved using functional RG.

\bibitem{lognote}
Weak additive logarithmic corrections may occur upon including higher-loop orders.
We found no evidence of those in the NRG calculations.

\bibitem{nrgrev}
R. Bulla, T. Costi, and T. Pruschke,
Rev. Mod. Phys. {\bf 80}, 395 (2008).

\bibitem{NRG_para}
The NRG parameters are $\Lambda=2$ and $N_s=800$.

\bibitem{hewson} A.~C.~Hewson,
{\em The Kondo Problem to Heavy Fermions}, Cambridge
University Press, Cambridge (1997).

\bibitem{tkdef}
For convenience, we have extracted $\TK$ from the NRG data by the crossover criterion
$T\chi_{\rm imp}(\TK)=0.1$.
For selected parameters, we have checked that this matches the frequently used definition via
the $T\to 0$ specific heat coefficient $\gamma_{\rm imp}$,
$\TK = 0.4128 (\pi^2/3) \gamma_{\rm imp}^{-1}$ \cite{hewson},
with a deviation less than 20\%.
Furthermore, NRG discretization effects lead to errors in $\TK$
of order 10\ldots 20\%.

\bibitem{dosnote}
We found the van-Hove singularities of the graphene DOS to cause numerical problems
upon conversion into a Wilson chain when $\mu\neq 0$. Therefore we employed a model DOS
with a broadened van-Hove peak, but otherwise the same characteristics as that of Fig.~2
of Ref.~\onlinecite{achim}. For this model DOS $\Jc = 4.7$\,eV while $\Jc = 4.3$\,eV for
the graphene DOS.

\bibitem{jconv}
In our notation of the Kondo Hamiltonian, $J$ is by a factor of 2 larger than that of
Ref.~\onlinecite{achim}.

\bibitem{flownote}
For an SU(4)-symmetric impurity, the fundamental p-h asymmetry of the critical fixed
point is expected to persist: the critical theory will have a structure similar to
Eq.~\eqref{aiminfu}, owing to the fact that the hybridization is marginal for $r=1$.

\bibitem{fanoth}
O. Ujsaghy, J. Kroha, L. Szunyogh, and A. Zawadowski,
Phys. Rev. Lett. {\bf 85}, 2557 (2000).

\end{thebibliography}
\end{document}